\documentclass[aps,prb,superscriptaddress,longbibliography,twocolumn]{revtex4-2}

\usepackage[en-US,useregional,showdow]{datetime2}
\DTMlangsetup[en-US]{ord=raise}

\usepackage{amsthm}
\theoremstyle{definition}

\theoremstyle{plain}

\usepackage{graphicx}
\usepackage{longtable}
\usepackage{wrapfig}
\usepackage{rotating}
\usepackage[normalem]{ulem}
\usepackage{amsmath}
\usepackage{amssymb}
\usepackage{capt-of}
\usepackage{hyperref}
\usepackage{amsthm}
\usepackage{xcolor}
\usepackage{soul}
\usepackage{physics}
\usepackage{slashed}
\usepackage{siunitx}
\usepackage{mathtools}
\usepackage{bm}
\usepackage{bbm}
\usepackage[capitalize]{cleveref}

\Crefname{equation}{Eq.}{Eqs.}

\usepackage[xparse]{tcolorbox}

\usepackage[inline]{enumitem}
\newlist{captionlist}{enumerate*}{2}
\setlist[captionlist,1]{label=\textbf{(\alph*)}}
\setlist[captionlist,2]{label=\textbf{(\alph{captionlisti}.\roman*)}}

\let\oldsout\sout
\renewcommand{\sout}[1]{{\color{red}{\oldsout{#1}}}}

\newcounter{para}
\newcommand{\para}{\par\refstepcounter{para}\textbf{{\color{cyan}[\thepara]}}\space}
\let\para\relax

\newcommand{\xCornell}{Department of Physics, Cornell University, Ithaca, NY, USA}
\newcommand{\xEwha}{Department of Physics, Ewha Womans University, Seoul, South Korea}
\newcommand{\xHarvard}{Department of Physics, Harvard University, Cambridge, MA, USA}
\newcommand{\xHarvardseas}{School of Engineering and Applied Sciences, Harvard University, Cambridge, MA, USA}
\newcommand{\GDM}{Google DeepMind, Mountain View, CA 94043}
\newcommand{\Google}{Google Research, Mountain View, CA 94043}

\begin{document}
\author{Haining Pan}
\affiliation{\xCornell}
\author{Nayantara Mudur}
\affiliation{\Google}
\affiliation{\xHarvard}
\author{Will Taranto}
\affiliation{\xCornell}
\author{Maria Tikhanovskaya}
\affiliation{\Google}
\affiliation{\xHarvard}
\author{Subhashini Venugopalan}
\affiliation{\Google}
\author{Yasaman Bahri}
\affiliation{\GDM}
\author{Michael P. Brenner}
\affiliation{\Google}
\affiliation{\xHarvard}
\affiliation{\xHarvardseas}

\author{Eun-Ah Kim}
\affiliation{\xCornell}
\affiliation{\Google}
\affiliation{\xEwha}

\title{Quantum Many-Body Physics Calculations with Large Language Models}
\begin{abstract}
Large language models (LLMs) have demonstrated an unprecedented ability to perform complex tasks in multiple domains, including mathematical and scientific reasoning. 
We demonstrate that with carefully designed prompts, LLMs can accurately carry out key calculations in research papers in theoretical physics. We focus on a broadly used approximation method in quantum physics: the
Hartree-Fock method, requiring an analytic multi-step calculation deriving approximate Hamiltonian and corresponding self-consistency equations. To carry out the calculations using LLMs, we design multi-step prompt templates that break down the analytic calculation into standardized steps with placeholders for problem-specific information. We evaluate GPT-4's performance in executing the calculation for 15  research papers from the past decade, demonstrating that, with correction of intermediate steps, it can correctly derive the final Hartree-Fock Hamiltonian in 13 cases and makes minor errors in 2 cases. Aggregating across all research papers, we find an average score of 87.5 (out of 100) on the execution of individual calculation steps. Overall, the requisite skill for doing these calculations is at the graduate level in quantum condensed matter theory. We further use LLMs to mitigate the two primary bottlenecks in this evaluation process: (i) extracting information from papers to fill in templates and (ii) automatic scoring of the calculation steps, demonstrating good results in both cases. The strong performance is the first step for developing algorithms that automatically explore theoretical hypotheses at an unprecedented scale.

\end{abstract}
\maketitle

\setcounter{para}{0}

\para The last few years have witnessed remarkable progress in developing large language models that can process and generate language, serving as powerful general problem solvers \cite{gpt3, moe, palm2, llama2}. These models have demonstrated impressive capabilities across diverse domains, including multidisciplinary benchmarks \cite{bigbench}, undergraduate-level math and science \cite{minerva, Wieman_physics}, coding \cite{codex_gpt, codingexp},  medicine \cite{medpalm, gpt4eval_med} and chemistry\cite{boiko2023emergent}. 
The developments lower barriers to entry between different domains of knowledge. Yet,
a major open question is whether it is possible to use LLMs to assist or augment human reasoning in specialized research settings such as theoretical physics, thereby pushing the generation of new knowledge. For example, the community has begun to explore whether LLMs can assist in solving problems in mathematics \cite{romera2023math}. The unique challenge of theoretical physics is that the domain requires inherently multi-faceted reasoning using language with specialized vocabulary, mathematics using symbols carrying specialized meanings, and code for numerical solutions. Developing an effective AI assistant will likely require going beyond scaling \cite{scaling_kaplan, chinchilla}.
 
\para 
Recent evaluations of Large Language Models (LLMs)\cite{gpt4paper,team2023gemini,huggingface_chatbot_arena} through qualitative analysis and quantitative benchmarks ~\cite{hendrycks2020measuring,medpalm,katz2023gpt} indicate that the best and most capable models have knowledge in a wide breadth of domains, much more so than the average person. This has led to an explosion of applications building on the potential of these models. In the natural sciences, recent works that demonstrate the utility of LLMs \footnote{We focus solely on LLMs available via model APIs and not on LLMs and foundational models trained or tuned on domain specific data} have focused on testing their knowledge in different domains \cite{ai4science,lala2023paperqa,MATHdataset, GSM8K, minerva,castro2023large,white2023assessment} or on their ability to generate code to help with automating experimentation processes that utilize software \cite{bran2023chemcrow,boiko2023emergent}. While having some domain knowledge is helpful, a model can be more useful if it can apply its knowledge to help solve a research problem. Strong examples of this are in the domains of mathematics and coding~\cite{lu2023chameleon,shen2023hugginggpt,alpha_geometry, romera2023math}. Less known is how theoretical scientists can harness LLMs' capabilities 
to augment their thinking process in scientific domains, requiring complex multi-step reasoning, where expertise (for humans)  takes time and specialization to acquire. 
While research is typically viewed as a creative activity, the extent to which large parts of the daily work of scientists can be automated is unclear. 
 A major component of theoretical physics research employs well-developed calculational frameworks, executing multi-step calculations integrating technical terminology with mathematical reasoning.  
Since state-of-the art LLMs typically use the arXiv \cite{taylor2022galactica} as part of their training corpus, it is plausible that LLMs have learned to perform sophisticated tasks in theoretical physics. A major challenge is to define the tasks and to establish an effective and robust framework for organizing tasks.

\begin{figure*}[ht]
  \centering
  \includegraphics[width=0.9\textwidth]{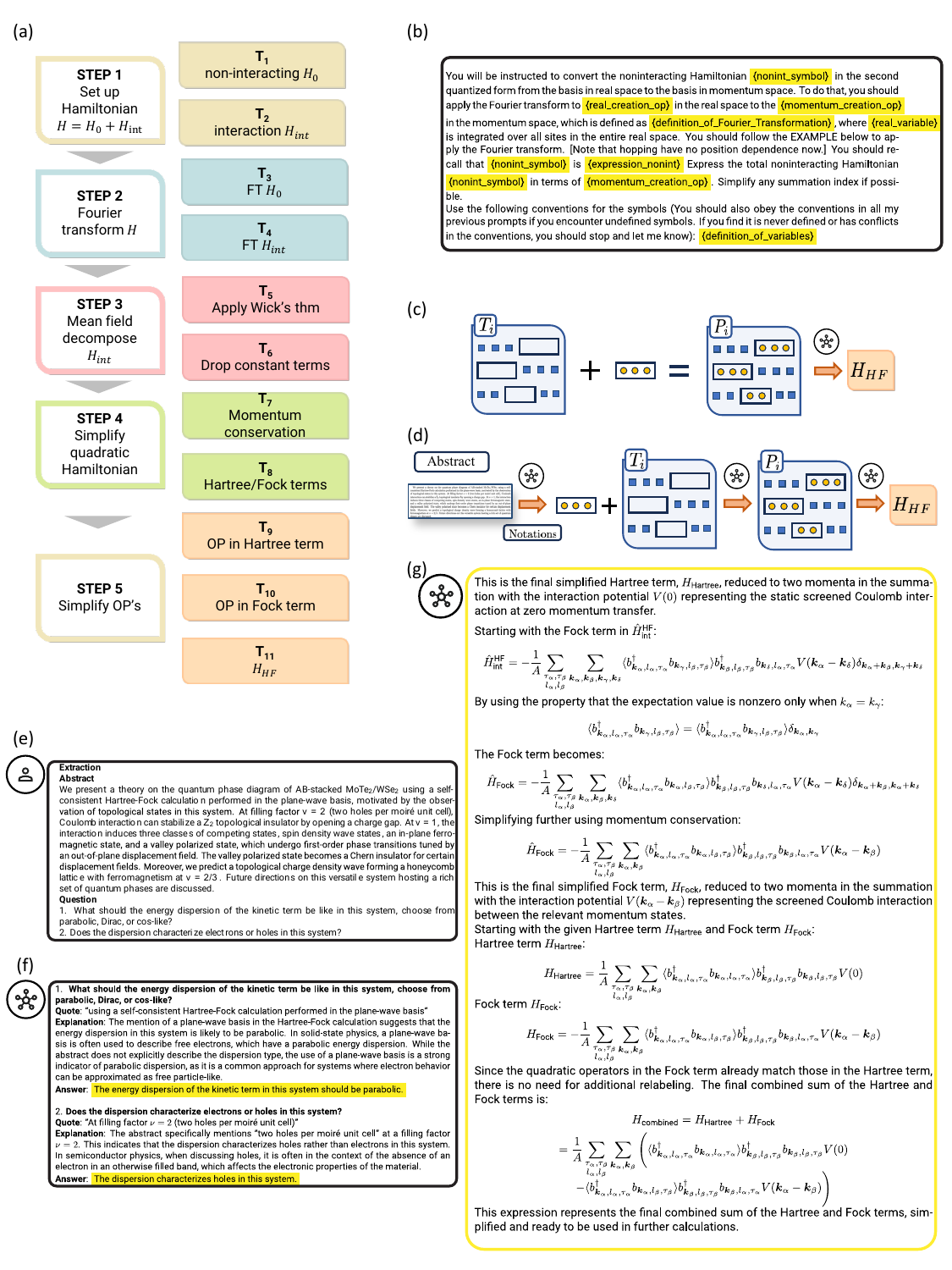}
  \caption{
    (a) The five conceptual steps of the derivation of the HF Hamiltonian and self-consistency equations and bite-sized tasks within each step. The HF template consists of the prompt template $T_i$ for each task $i$.     
    (b) An example template {$T_3$}. The placeholders are highlighted. We turn the template into a prompt for the task $3$ by specifying the placeholders for the given paper in the database. 
    (c) The schematic for generating the prompts from the template with placeholders (empty boxes) using human-supplied information (boxes with dots). 
    (d) The schematic for generating the prompts from an abstract. We give an abstract to a LLM and query the LLM to infer system specific information from the abstract and fill relevant placeholders in the template. Since notations are not specified in the abstract, we supply placeholders corresponding to the notations. The combination is a complete prompt. 
    (e) An example of a query asking an LLM to infer system specific information. 
    (f) An example response from GPT-4 to the query of panel (e). We required the response to consist of the quote, an explanation, and the answer. The answers are highlighted. 
    (g) An example of a response to the final prompt by GPT-4 for Ref.~\cite{pan2022topological} corresponding to $T_{11}$.
      }
   \label{fig:prototype-problem}
\end{figure*}

\para In this work, we investigate the ability of GPT-4 to assist theoretical research in quantum many-body physics. To our knowledge, this is the first investigation in evaluating LLMs on research-level problems in physics. We focus on a well-established and widely used calculational framework of the Hartree-Fock (HF) mean field theory \cite{altland_simons}. These Hartree-Fock methods follow the treatment employed in condensed matter physics to study effective theories of quantum many-body systems, in contrast to first-principles approaches for molecules and solids.
Although such calculations are widely used tools in theoretical research \footnote{Over 6456 papers mention Hartree-Fock in the abstract of papers in the cond-mat arXiv preprint server over the last decade.}, learning to do the calculations reliably requires years of study. As a first step towards automating HF calculations, we built a multi-step prompt template: the HF template (see Figure~\ref{fig:prototype-problem}(a)). This template consists of step-by-step directions suitable for instructing beginning graduate students to do the calculations.
The template has placeholders to set up notation and enter problem-specific information Fig.~\ref{fig:prototype-problem}(b-c). As an evaluation testbed, we curated a corpus of 15 recent research papers reporting results of HF mean-field theory for various distinct physical systems.
Given the correctly filled-in prompt templates, we demonstrate that GPT-4 can accurately carry out the calculations. More ambitiously,
  GPT-4 can sometimes extract or infer the necessary problem-specific information just from an abstract, supplying the placeholders and carrying out the calculation (see Fig.~\ref{fig:prototype-problem}(d-g)). 
Finally, 
we test GPT-4's ability to extract correct placeholders from human-selected excerpts of the papers, demonstrating the ability to read critical information.  

\begin{figure*}[ht]
  \centering
  \includegraphics[width=\textwidth]{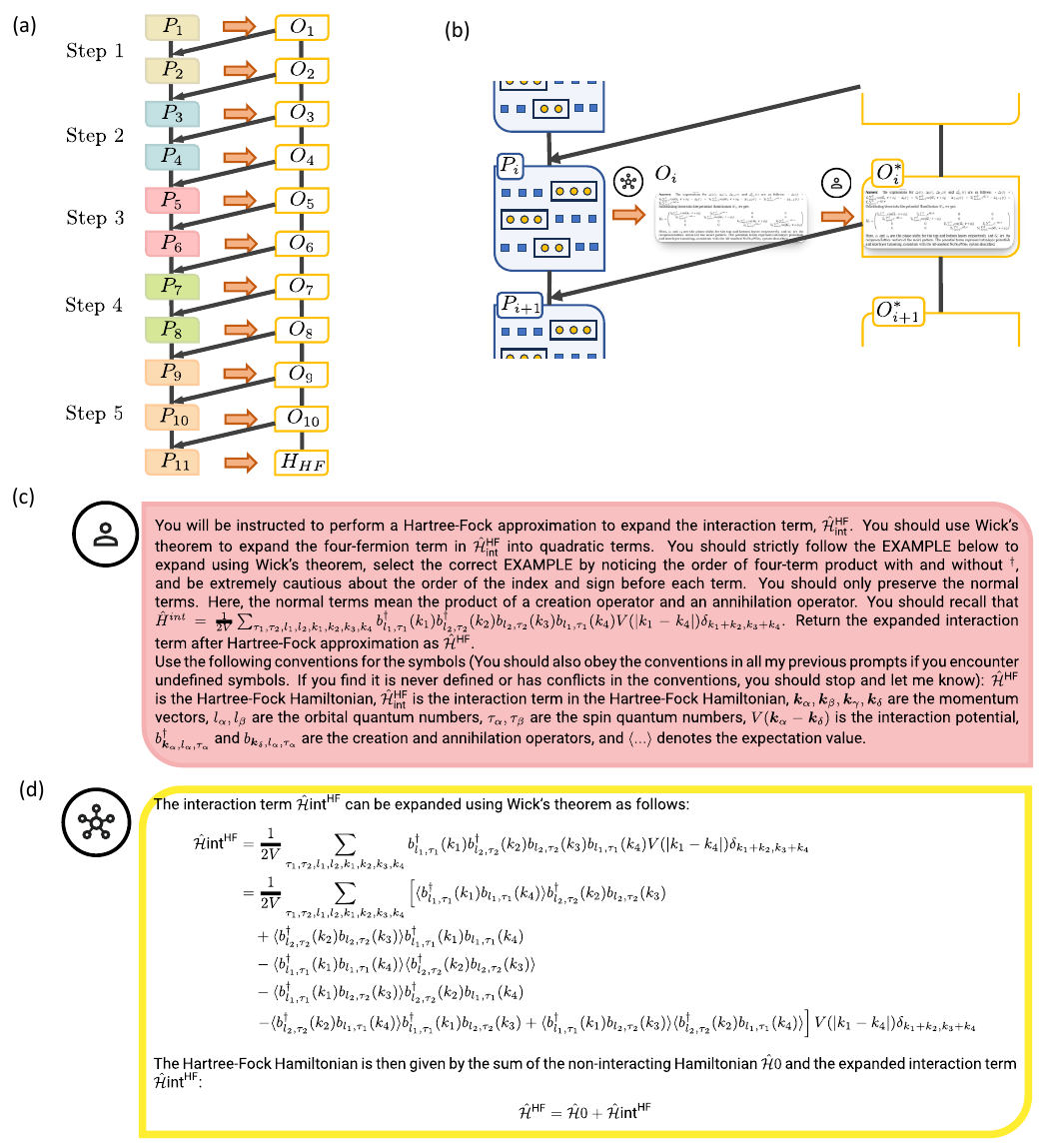}
  \caption{
  (a) The execution workflow using the full prompt set based on the HF template. Each prompt builds on the outputs of all the previous steps. Specifically, the prompt for the task $i$, $P_i$ incorporates the corrected output $O_{i-1}$ of the previous prompt. 
  (b) The schematic of evaluation and correction for each task $i$. Each output $O_i$ to the prompt $P_i$ executing the task $i$ is evaluated by the human evaluator and corrected, if necessary. The verified output $O_i^*$ is incorporated into the next prompt $P_i$.
(c) An example of the prompt $P_5$ for reproducing the calculations in Ref.~\cite{pan2022topological}. 
(d) An example of the execution outcome $O_5$. This output is correct, hence correction was not necessary and \(O_i=O_i^*\).
  }
   \label{fig:template}
\end{figure*}

\para The HF method is an instance of mean-field theory, where interactions are replaced by ``mean fields'' proportional to an order parameter, whose fluctuations are ignored.
The application of the method neatly separates into an analytical stage of deriving the HF Hamiltonian $H_{HF}$ and the associated order parameter $\langle\hat{\Delta}_{\rm symm}\rangle$ and a computational stage of iteratively solving the self-consistency equation. 
Deriving $H_{HF}$ for a given problem involves a series of steps broken down in Fig.~\ref{fig:template}(a),  requiring synthesizing natural language concepts in mathematical terms with symbols and equations.  
The first step (STEP 1) establishes the dynamical degrees of freedom and fixes the single-particle Hilbert space. The non-interacting Hamiltonian $H_0$ specifies the particle flavors (spin, orbital, valley, layer, etc.) and the dispersion and potential specific to each flavor. The interaction Hamiltonian $H_{\rm int}$ specifies the distance and flavor-dependent interaction. The system Hamiltonian $H$ should reflect the symmetries of the problem, which will determine the block-diagonal structure of the ultimate $H_{HF}$. The Fourier transforms in STEP 2 determine the momentum dependence of the eventual $H_{HF}$. In STEP 3, Wick's theorem is applied to decompose the Hamiltonian based on mean-fields. In STEP 4, the quadratic Hamiltonian is simplified and organized into Hartree terms diagonal in spin space and Fock terms off-diagonal in spin space. In STEP 5, the symmetries of the system are used to reveal order parameter structures $\langle\hat{\Delta}_{\rm symm}\rangle$ in the Hartree and Fock terms.
Once $H_{HF}$ is obtained for all symmetry-breaking channels, 
researchers choose the channel of interest and numerically solve the self-consistency equation (see SM Section A~\cite{SM}). 
With the correct
$H_{HF}$ in hand, the computational cost of solving the resulting self-consistency equation is negligible. 
Beyond setting up the matrix self-consistency equations for the numerical evaluation, the
analytical derivation of $H_{HF}$ offers physical insight into the problem by 
revealing the role of interactions in different symmetry-breaking sectors for the system.

\para Fig.~\ref{fig:prototype-problem}(c) shows a schematic of an automated derivation of $H_{HF}$ starting from an abstract of a paper. The idea is to supply a synopsis of a calculation in natural language and ask an LLM to figure out the problem-specific information from the synopsis by answering 10 questions (see SM Section E~\cite{SM}). With each answered question, we ask the LLM to complete the corresponding placeholders in the HF template. We then prompt the LLM with the template to derive  $H_{HF}$ for the problem of interest. We demonstrate the feasibility of the idea using the abstract of arXiv:2111.01152~\cite{pan2022topological} and GPT-4.  
Fig.~\ref{fig:prototype-problem}(d) shows a portion of the prompt where we supply the abstract and ask questions. As we show in detail in \cite{SM}, we instruct  GPT-4 to quote the part of the abstract on which it is basing its answer and explain its reasoning before committing to an answer. This structure led to GPT-4 correctly answering all of the 10 questions as exemplified in Fig.~\ref{fig:prototype-problem}(e). While most questions were correctly answered on the first try, some questions requiring an inference using prior knowledge took 2-3 attempts for GPT-4 to arrive at correct answers (see SM Section E~\cite{SM}) As shown in Fig.~\ref{fig:prototype-problem}(g), after a six page calculation, GPT-4 outputs the correct $H_{HF}$ compared to Eq.(S9) in arXiv:2111.01152~\cite{SM} given the prompts with correct placeholders. While the performance of an LLM on this ambitious task naturally depends on the degrees of specificity given in the abstract, this test case establishes that the vision of giving a synopsis of a calculation in natural language to an LLM to carry out a Hartree-Fock calculation is within reach. 

\para To build general prompt templates for deriving $H_{HF}$, we break down the process into five high-level steps, as shown in Fig.~\ref{fig:prototype-problem}(a).
The prompt templates further break down the steps into natural algorithmic pieces $T_i \, (i=1,\cdots, 11)$ with placeholders for problem-specific information (see Fig.~\ref{fig:prototype-problem}(b) and SM Section C~\cite{SM}). To reduce model hallucination, we give examples within the templates, just as we would to teach a beginning graduate student.
We consider three general settings of interacting fermion problems: whether or not the system is analyzed in the continuum limit, and whether the Hamiltonian is expressed in second-quantized form (see SM Section C~\cite{SM}).
Once the placeholders are specified (see Fig.~\ref{fig:prototype-problem}(b)), we use a sequential series of prompts $\{P_i\}$ to elicit the LLM to perform the calculation. The prompts instruct the LLM to use the information specified in previous steps, as well as the outputs from previous steps (see Fig~\ref{fig:template}(a)). 
To avoid propagating errors, the response to each prompt is evaluated and (potentially) corrected before becoming the grounds for the next prompt (Fig~\ref{fig:template}(b)).
Fig.~\ref{fig:template}(c,d) give an example of a prompt-response pair applying Wick's theorem in step 3.  
To evaluate the efficacy of the HF-template, we tested the template on 15 papers selected from 807 preprints published in American Physical Society journals during the last decade.  The precise rules we used to select papers, the complete database of papers we have analyzed, templates, and complete codes designed for multiple scenarios are available in SM section C in~\cite{SM}. In the remainder of the paper, recent checkpoints of GPT-4 from mid-2023 were used for all evaluations \footnote{Evaluations were carried out using checkpoints `gpt-4' and `gpt-4-0613' referenced in https://platform.openai.com/docs/models/gpt-4-and-gpt-4-turbo. At the time that the experiments in the paper were performed, `gpt-4' pointed to `gpt-4-0613'. The abstract to execution experiment was performed using GPT-4 queried via the web interface.}.

\begin{figure*}[ht]
  \centering
  \includegraphics[width=\textwidth]{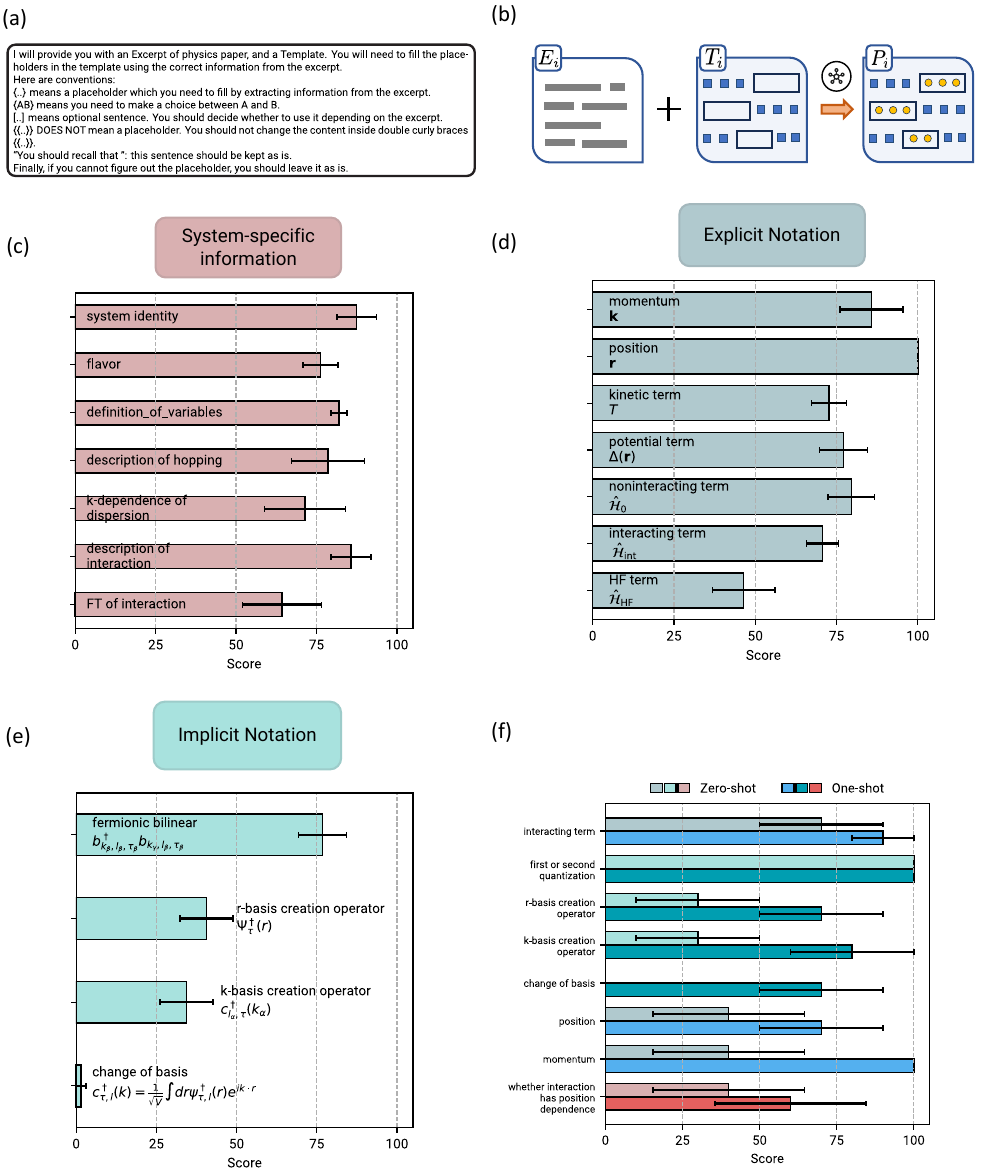}
  \caption{
  (a) The extraction prompt. This prompt is supplied to GPT-4 together with an excerpt $E_i$ and the HF template $T_
i$. The prompt instructs the LLM to locate the placeholders in the template and replace the dummy labels with information it extracts from the excerpt. The output will be an execution prompt $P_i$. 
(b) Schematic of excerpt-based information extraction using the prompt in (a). 
(c-e) Mean and standard error on the mean of the score for placeholder completion for a subset of placeholders, organized by the type of information associated with the placeholders: Information specific to the nature of the system (c); notation explicitly present in the excerpt (d); notation that needs to be inferred (e). 
(f) Comparison of the extraction in zero-shot and one-shot scenarios, using the mean performance over five papers to define the bar's length and the standard error of the mean for the error bar. 
 }
  \label{fig:extraction}
\end{figure*}

\para In evaluating the HF-template, a major bottleneck was substituting relevant paper-specific information into over 76 placeholders of the HF-template for each paper. This template completion requires one to synthesize the conventions and notation of quantum many-body physics with problem-specific information for each paper and infer the relevant information from the context and the label for the placeholder. 
Hence the template completion task defines an advanced and purposeful information extraction task requiring technical expertise. We explored automation of prompt completion
both as a stepping stone towards full automation and as a gauge of the LLM's ability to reason. Specifically, we used a single universal prompt shown in Fig.~\ref{fig:extraction} to turn each template step $T_i$ into a completed paper-specific prompt $P_i$ by extracting and inferring information from the human-selected excerpt $E_i$ (see Fig.~\ref{fig:extraction}(b)). 

\para  We evaluated information extraction for each placeholder by averaging the scores from two authors (Y.B. and H.P.) who have significant experience in quantum many-body physics. Each placeholder is assigned a categorical score of either 0 (no credit), 50 (partial credit), and 100 (full credit). To glean insight into the score distribution, we grouped the placeholders into three categories:  
 {\it system-specific information}, {\it explicit notation} (notation commonly specified in the papers), and {\it implicit notation}. The latter refers to conventions in the field that are often unspecified in the papers. 
 Fig.~\ref{fig:extraction}(c-e) shows the scores for a subset of placeholders in each category.
 (see SM Section F~\cite{SM} for the complete set.) 
GPT-4 did a remarkable job at extracting system-specific information from excerpts as shown in Fig.~\ref{fig:extraction}(c), consistent with its success in extracting and inferring such information from the abstract for the preprint {arXiv:2111.01152}~\cite{pan2022topological} (Fig.~\ref{fig:prototype-problem}(c-f)). GPT-4 also consistently performed well when extracting 
the notation and convention appearing in the excerpt (see Fig.~\ref{fig:extraction}(d)). 
Where it struggled the most was in specifying
placeholders which required synthesizing prior knowledge of quantum many-body physics conventions with 
what could be inferred from the template and the excerpt to introduce a suitable notation (see Fig.~\ref{fig:extraction}(e)).
An example of the ``change of basis'' that GPT-4 failed  is to fill 
 the placeholder ``definition\_of\_Fourier\_Transformation"
 in Fig.~\ref{fig:prototype-problem}(b) with a correct equation using the second-quantized operators. To succeed in this task, the LLM needs prior knowledge of the second-quantized operators in position and momentum basis. Moreover, the LLM has to infer the task from the label and context of the placeholder.  

\para To explore how to improve the information extraction performance, we examine the impact of a one-shot evaluation prompt \cite{gpt3} on the extraction task for the template $T_4$, Fourier transforming the interaction term $H_{\rm int}$ (Fig.~\ref{fig:prototype-problem}(a)). Here, ``one-shot evaluation" means we provide a single example of an excerpt and its correct target output in the prompt (see SM Section H~\cite{SM}). The initial performance of GPT-4 on the 40 placeholders across 5 papers for $T_4$ was $44\pm 8$. 
 Fig.~\ref{fig:extraction}(f) shows that the one-shot prompt 
significantly enhances the performance,  more than doubling the mean performance. The performance across all placeholders in the one-shot setting was $80\pm 6$. This implies information extraction for the auto-generation of prompts may be achievable with the help of more exposure to requisite knowledge and examples.

\begin{figure*}[ht]
  \centering
  \includegraphics[width=\textwidth]{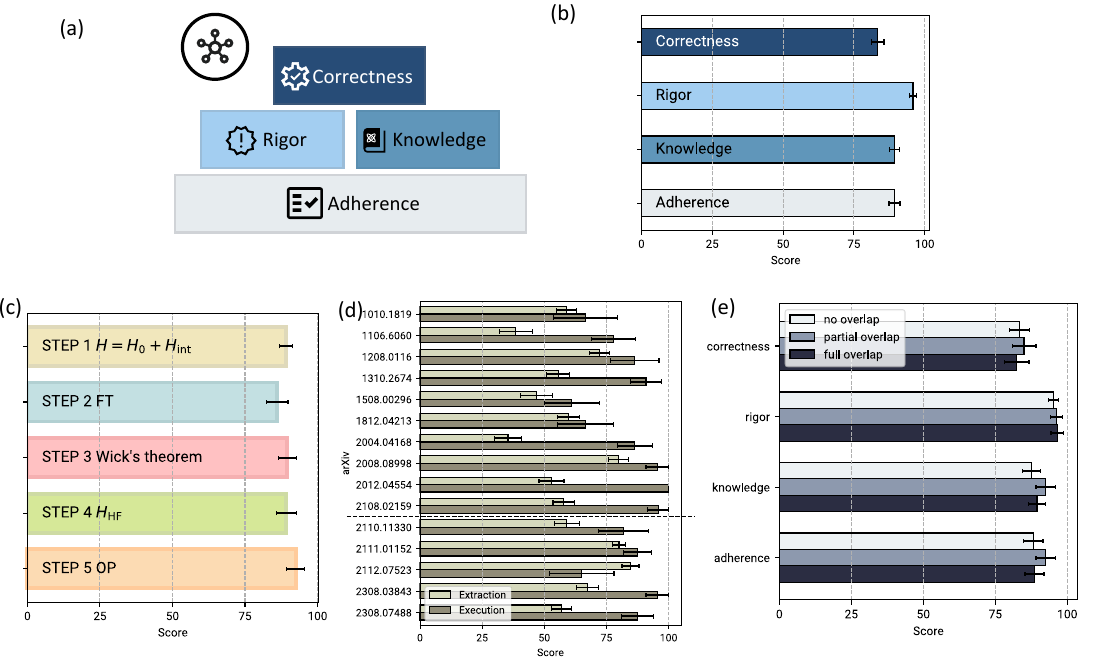}
  \caption{
  (a) The four-layered rubric system for evaluating an LLM's output $O_i$ in response to each prompt $P_i$.  {\bf Adherence}: how closely the LLM adheres to the instructions. {\bf Rigor}: how accurate is the mathematical derivation. {\bf Knowledge}: how consistent is the LLM's reasoning with the laws of physics. {\bf Correctness}: how correct is the LLM's response. 
(b) The rubric-dependence of the performance. The average score for each rubric layer across all outputs for all papers and their standard deviations.
(c) The task-dependence of the performance. We averaged the score for each prompt across the four rubric layers. Then these average scores were averaged over the prompts belonging to each step of the derivation as broken down in Fig.~\ref{fig:template}(a). 
(d) The paper-dependence of the performance on information extraction and execution. The average score across all the placeholders for a given paper over the excerpt-based information extraction detailed in Fig.~\ref{fig:extraction} is shown in the lighter sage green. The average score across all rubric layers and prompts for deriving the $H_{HF}$ for a given paper is shown in darker olive green. For both extraction and execution, the error bars were calculated by averaging over all papers and placeholders/tasks. The dashed line between arXiv:2108.02159 and arXiv:2110.11330 marks the separation between papers
before and after the training data cutoff date.
(e) The dependence of the execution score, for each rubric layer,  on the degree of the overlap between the correct output $O_i^*$ and the text of the target research paper.  
  }
   \label{fig:execution}
\end{figure*}

\para
Finally, we turn to the main focus of this paper: evaluation, across all steps and papers, of the full HF calculation using our correctly filled-in templates (see SM Section C~\cite{SM}). The LLM responses are then scored by an expert in Hartree-Fock calculations (H.P.). The response to each prompt is a paragraph of varying length depending on each problem (see Fig.~\ref{fig:template}(d) for example). In order to evaluate the execution responses in a fine-grained manner that is nevertheless standardized across all the papers, we introduced a four-layered rubric system (Fig.~\ref{fig:execution}(a)).
The four rubric layers are (1) how closely the LLM executes the instructions ({\bf Adherence}); (2) accuracy in the LLM's mathematical derivations ({\bf Rigor}); (3) consistency in the LLM's reasoning with the laws of physics  ({\bf Knowledge}); and (4) correctness in the LLM's final answer ({\bf Correctness}). Each response is scored from all four layers on a categorical scale of 0 (no credit), 50 (partial credit), and 100 (full credit). The evaluation results across all papers presented in Fig.~\ref{fig:execution}(b) shows that our HF-template based approach to the Hartree-Fock mean field theory works well at all layers. For {\bf Correctness}, the score of 100 requires the output to be exactly correct. Often, the results of intermediate steps did not appear in the paper and were calculated by the scorer. For {\bf Rigor},  occasional errors in indices or subscripts result in partial credit. For example, we give partial credit for {\bf Knowledge} if the LLM makes a sign error in the momentum shift needed due to the presence of a magnetic field.
For {\bf Adherence}, for example, partial credit would be awarded if the LLM insists on using conventions different from that explicitly directed. We note that the {\bf Correctness} score is the lowest since errors in any of the three layers will reduce this score. The high and reliable score of above 95 in {\bf Rigor} shows
that GPT-4 was quite capable of carrying out the calculations correctly. In fact, we discovered typos in research papers in our corpus in the process (see SM Section G~\cite{SM}).  
Aggregating across all rubric layers and all papers, the average score of GPT-4 is 87.5 from Fig.~\ref{fig:execution}(d), with slight variation across the papers. This is a remarkably high score for a generically trained LLM, signifying expert-level performance.

\para  Fig.~\ref{fig:execution}(c) compares the score across the five steps of the prompt template (Fig.~\ref{fig:prototype-problem}(a)), averaged over all four rubrics layers. The performance is uniformly high across the steps, demonstrating that the fine-grained approach of the HF template allows the model to succeed in system-specific Hamiltonian building (STEP 1) and quantum many-body operator algebra in the mean-field decomposition (STEP 3). This uniformly high score across the entire corpus of papers in executing the steps of the calculation is in striking contrast to relatively irregular and overall lower scores in extracting information for the placeholders as summarized in Fig.~\ref{fig:execution}(d). The contrast reveals that even when the model can carry out the actions of a specific analytic calculation following a natural language instruction, making specific plans for the action through an informed reading framed through a calculation structure can be challenging.

\para 
A challenge with evaluating LLMs -- in particular GPT-4, for which the full training dataset details have not been published -- is to what extent the evaluation set overlaps with the training set, and whether there is evidence for generalization to new, never-before-seen problems. Unlike other evaluation tasks, which are typically algorithmically curated or generated, we seek to evaluate LLMs on a research task. Hence, generating new problems is considerably more involved in our setting. Instead, we investigate the issue of generalization through two indirect means. (1) The GPT-4 model used in this paper has the training data cutoff date of 
September 2021, marked as a dashed line in our chronological summary of overall scores in Fig.~\ref{fig:execution}(d). While OpenAI reports that there are small amounts of more recent data used \cite{gpt4paper}, the fact that the cutoff date does not affect the score is encouraging. (2) A novel aspect of our task curation process is that it required us to fill in the intermediate steps of the calculations laid out in the HF-template (Fig 1(a)); they are not explicitly presented in the papers.
We divide our execution steps across the corpus of papers into three categories depending on the degree to which the results of the step appeared explicitly in the paper of interest, and plot the scores for each prompt response averaged over all the tasks in each category. As shown in Fig.~\ref{fig:execution}(e), the scores exhibit a flat distribution conditional on this categorization. These two observations support the conclusion that LLMs can do non-trivial aspects of HF calculations, regardless of whether the solution explicitly appears in its training data.

\para Finally, to investigate whether an LLM could further be used to supplement human evaluation, we examined the performance of GPT-4 on scoring its (own) execution responses as \textit{Incorrect} or \textit{Correct} in a zero-shot setting as well as a setting where the model is prompted to return a rationale (see SM Section I~\cite{SM}). We computed the class-balanced accuracy of the `LLM-Scorer', where the two classes are \textit{Incorrect} and \textit{Correct}, against the human-assigned \textbf{Correctness} scores. The `LLM-Scorer' had a class-balanced accuracy of 69\% and 74\% in the zero-shot and the few-shot-with-rationale experiments, respectively. In the setting where the model is prompted to return a rationale for its score, the `LLM-Scorer' was able to identify 72.5\% of problems that had a human-assigned \textbf{Correctness} score of 0 or 50 as incorrect. We find the agreement between LLM self-scoring with human scoring to be promising; developing automated scoring procedures would help mitigate the human-evaluation bottleneck and enable scalable evaluation and improvement of models in the future.

\para 
In this work, we investigated LLMs' abilities to execute HF mean-field theory, a core tool in quantum many-body physics, by creating an evaluation corpus from research papers in the field. To do so, we designed a collection of general-purpose prompt templates to break the task into smaller but natural calculation steps. We also experimented with using LLM to fill in the templates for individual research papers. In executing the steps of HF mean-field theory, GPT-4 scored 87.5 out of 100 on average, aggregated over all steps and all research papers in our corpus, despite the tasks requiring graduate-level knowledge of quantum many-body physics. To our knowledge, this is the first evaluation of LLM abilities to execute a core component of research-level physics, used in numerous papers in condensed matter theory. Our work demonstrates that a broadly employed theoretical tool for scientific research can take advantage of existing LLM capabilities.

\para 
The analytic derivation of the HF Hamiltonian and associated self-consistency equations is a common first pass at a complex problem of quantum many-body physics when new systems and phenomena are discovered. 
Hence the HF template we developed for carrying out the derivation using LLMs
has the potential to aid accelerated exploration. Moreover, the template and the evaluation scheme we introduced can also serve as a testbed for evaluating future improvements on LLM reasoning capabilities.  Myriad improvements can be explored, from fine-tuning an LLM to achieve specific domain knowledge in HF calculations, to more elaborate examples in each step of the prompt template, to allowing the LLM to call computational tools. Ultimately, these will allow researchers to explore the rich and complex phase space of possibilities in modern material systems efficiently. 
We anticipate, given the good baseline performance of GPT-4 on each part of this process, that these different tasks could potentially be bootstrapped towards better models in an automated manner. Looking farther ahead, augmenting LLMs with the ability to call computational tools would enable seamless translation between language, symbolic mathematics, physical insight, and numerical solvers, a combination with potentially powerful consequences for scientific research. 

\para
HP acknowledges support by the National Science Foundation (Platform for the Accelerated Realization, Analysis, and Discovery of Interface Materials (PARADIM)) under Cooperative Agreement No. DMR-2039380. WT, E-AK were supported by  OAC-2118310 (09/15/2022 to 08/31/2027 ), HDR Institute: The Quantum Institute for Data and Emergence at Atomic Scales (Qu-IDEAS).  This research is funded  in part by the Gordon and Betty Moore Foundation’s EPiQS Initiative, Grant GBMF10436 to E-AK.


\newpage
\bibliographystyle{naturemag}
\bibliography{refs}

\end{document}